Estimation of black bear abundance by management zone in Wisconsin

Maximilian L. Allen[1], Nathan M. Roberts[2], Andrew S. Norton[3], Timothy R. Van Deelen[4]

[1] *Illinois Natural History Survey, University of Illinois, 1816 S. Oak Street, Champaign, IL 61820, USA*

[2] *Wisconsin Department of Natural Resources, 107 Sutliff Avenue, Rhinelander, WI 54501, USA*

[3] *South Dakota Game, Fish and Parks, 4130 Adventure Trail, Rapid City, SD 57702*

[4] *Department of Forest and Wildlife Ecology, University of Wisconsin, 1630 Linden Drive, Madison, WI 53706*

[1]*Email*: maxallen@illinois.edu

**ABSTRACT**

Estimates of population abundance are fundamental to wildlife management and conservation, but are difficult to obtain across large geographic scales and for cryptic species. We used a state-space model with age-at-harvest data in a Bayesian framework to model American black bear (*Ursus americanus*) abundance and demographic parameters in four management zones in Wisconsin from 2011-2017. We had limited demographic data available from the population, and relied upon a) the model, b) age-at-harvest data, and c) informative prior distributions from a literature review. The estimated posterior means and distributions for abundance and demographic parameters from our models were reasonable for each management zone, and indicated a decreasing trend in zones A and B, and a generally stable trend in zones C and D. The age-at-harvest data updated the posterior distribution and means for initial population size, harvest season survival, and non-harvest season survival, with a notable increase in precision for the survival values. A strength of the model for managers is the formalized process for providing biologically supported information as prior distributions to transparently accommodate expert opinion with measures of confidence when estimating wildlife populations, which can then be updated by the age-at-harvest data and model structure. The integration of prior information and age-at-harvest state-space models in a Bayesian framework efficiently leverage all available information for making zone-specific abundance estimates for the management of harvested species. This may create more informative data for decision-makers when setting harvest quotas, and could lead to more effective monitoring, conservation, and management of cryptic carnivore species.

## Introduction

Estimates of population abundance are fundamental to wildlife management and conservation (Leopold 1986, Skalski et al. 2005), but estimates are difficult and expensive to obtain across large geographic scales (Skalski et al. 2005). This is especially true for carnivores, which are cryptic and occur at low-densities (Karanth and Nichols 1998, Allen et al. 2016a). Consequently, carnivore managers often base population estimates on extrapolations from small data sets, which makes it important to use appropriate sampling design and statistically rigorous models and obtain accurate estimates (e.g., Skalski et al., 2005). For management of hunted populations, using models based on age-at-harvest (AAH) data to create population estimates is often most practical because AAH data are collected across management zones, and the cost of obtaining the age and sex of harvested animals is minimal compared to other data collection methods (Skalski et al. 2005, Norton 2015, Allen et al. 2018a).

Several statistical models have been developed that use AAH to accurately estimate population abundances (Conn et al. 2008, Fieberg et al. 2010, Skalski et al. 2011, Norton 2015). To summarize, we distinguish between these based on whether the past models used were fit in a frequentist or Bayesian framework and how the AAH data are modeled. Past frequentist approaches have included statistical population reconstruction (SPR) that integrated hunter effort (e.g., Skalski et al. 2011), and state-space models that integrated food availability (e.g., Fieberg et al. 2010). Past Bayesian approaches have included data augmentation methods that are similar to SPR and integrated tag recovery (e.g., Conn et al. 2008), and state-space models that integrated known-fate survival data (e.g., Norton 2015) or produced reasonable estimates without integrating auxiliary data (Allen et al. 2018a). Bayesian approaches allow the analyst to provide biologically supported values and constraints on model parameters as prior distributions, creating a formal process that transparently accommodates expert opinion when estimating wildlife populations. This formal process also allows for updating our knowledge and understanding of a given population by formally introducing knowledge of the system as prior values/distributions. Our knowledge of the system (the prior values/distributions) are then updated with the information in the new data (in this case the AAH data from harvest data) and summarized as posterior distributions that fully expresses our new understanding of the system. State-space models allow the user a flexible framework to completely describe the temporal resolution of a biological process (e.g., survival or reproduction) that is only partially observed on a recurrent basis (e.g., annual age-at-harvest data). However, reliability of inference is still contingent on the resolution and quality of the observed data and prior models, which requires the modeler to completely understand limitations of the data.

American black bears (*Ursus americanus*) are a spatially dispersed (Rogers 1987, Taylor et al. 2015) and long-lived solitary carnivore (up to 31 years in the wild; Allen *et al.*, 2017). They have a relatively low reproductive rate and heavily invest in parental care (Rogers 1987), frequently not reaching sexual maturity until they are 2-7 years old (Rogers 1987, Noyce and Garshelis 1994, Allen et al. 2017). Consequently, as a K-selected species, black bears have slow population growth rates (Pianka 1970) and are susceptible to over-harvest (Garshelis and Hristienko 2006).

In Wisconsin, black bears are a big game animal, that is managed by the Wisconsin Department of Natural Resources (WDNR), and whose populations and harvests have been growing over the last 45 years (MacFarland 2009, Sadeghpour and Ginnett 2011, Allen et al. 2018a). The WDNR has estimated the population abundances of black bears in 4 management zones over the last 3 decades using a deterministic accounting-style model. The model is dependent on initial population size, incorporates fixed estimates of demographic parameters, and does not report any measure of variance for the estimates (Allen et al. 2018). The model also does not always estimate populations accurately compared to independent estimates (MacFarland 2009). Independent population estimates from tetracycline-bait mark-recapture studies (see MacFarland 2009) have been integrated into the WDNR population model to increase accuracy, but these independent estimates are expensive and often conducted years apart, and do not fix the underlying problems of the model not estimating variance or allowing variation in demographic parameters. Allen et al. (2018) developed a state-space model using AAH in a Bayesian framework for black bears that incorporates variation in population and harvest demographics over time which could be used to update the population estimates for each management zone in Wisconsin.

State-space models may be ideal for estimating wildlife populations (Norton 2015, Allen et al. 2018a), but have been infrequently used by wildlife managers to date. Our goal was to evaluate whether AAH state-space models, based on Norton (2015) and Allen et al. (2018), can be used to effectively estimate zone-specific black bear abundance and demographic parameters in Wisconsin for use in management (setting harvest objectives for recreation and conservation). Our objectives were to 1) use the AAH state-space models developed by Allen et al. (2018) for black bears to create zone-specific estimates, and 2) compare the zone-specific posterior distributions of abundance to determine if the model can predict different trends based on the given zone. 3) Compare the zone-specific prior and posterior distributions of each demographic parameters to determine how they were updated by the model.

## Methods

*Study Area*

The WDNR manages bear populations in 4 distinct zones, with each zone having unique quotas and hunting regulations (MacFarland 2009). Most of the bear population resides in the northern half of Wisconsin, which is represented by management zones A, B, and D. These zones are characterized by northern mixed forest ecotone. Zone C has seen expansion of bear range in recent years, with the population being most abundant in the northwest part of the zone, and absent in the southeast part of the zone.

In Wisconsin, the bear hunting season was open for 35 days during each year of our study, with the season beginning the first Wednesday after Labor Day each year. The annual harvest quotas for each management zone are set each year. The harvest of black bear cubs or females with dependent cubs is prohibited in Wisconsin. Hunting with trained hounds is allowed in management zones A, B, and D but not in Zone C.

*Model Parameters and Processes*

We used the model created by Allen et al. (2018) to estimate abundance ($N$) of the black bear population in each management zone in Wisconsin immediately prior to the hunting season. We used 6 years of black bear harvest data collected from mandatory registration in Wisconsin to determine the total annual observed harvest ($O$), and the number of harvested animals with known age and sex ($C$). A tooth was extracted from each harvested black bear to age through counting cementum annuli. We annotate population size as $N_{a,s,y}$, where $a$ = age, $s$ = sex, and $y$ = year, for the indicated population size.

We used the model created by Allen et al. (2018), including most of the reasonably informative prior distributions, and allowing annual stochasticity in the harvest season survival but not for other demographic parameters. We changed the model created by Allen et al. (2018) by adjusting A) the initial population sizes, B) the proportion of bears in the population in each sex class, and C) only used 6 years (2011-2016) of data to match the data from our initial population size estimates. We based our initial adult population sizes for each zone on independent capture-recapture estimates generated from tetracycline marking from 2011 ($N_A$ = 8,478, $N_B$ = 6,075, $N_C$ = 7,748, and $N_D$ = 8,633; Rolley and Macfarland 2014). To account for known underestimates in the proportion of females and overestimates in the proportion of males in initial population estimates in the model of Allen et al. (2018), likely due to male-biased harvest selection (Malcolm and Van Deelen 2010), we considered 55% of the population to be female proportion and 45% to be male. To estimate the proportions of $N$ in each age class in the initial population size, we set abundance in age class 1.5 years old based on the values from the tetracycline marking and assumed survival for each age transition from 1.5 to 10.5+ was equal to our mean prior value. We assumed cub survival was equal for male and females, and calculated the abundance of the 10.5+ year old age class to be three times greater than the 10.5 age class.

Our state-space models consisted of two process models whose likelihoods were jointly modeled (Buckland et al. 2004, Norton 2015). The population process model (two-sex, ten-stage population projection matrix [Caswell 2001]) was based on the unobserved/latent population matrix, while the observation state process used the harvest data from Wisconsin in each given zone (Newman et al. 2009, Norton 2015). We achieved regularization of parameter estimates by using a hierarchical structure for parameters to smooth annual or age-cohort variation to mean values for each modeled parameter based on information about black bear ecology from the literature review.

We fit our models using Markov Chain Monte Carlo (MCMC) methods and report the posterior means and distributions for each parameter from the MCMC samples. We created zone-specific state-space population models for each of the four management zones in Wisconsin to estimate the black bear population using actual harvest data from 2011–2016 and our prior distributions. We fit our models in Program R (R Core Team 2017) using *JAGS* (Plummer 2003) and the R package *rjags* (Plummer 2006). In each model we used 220,000 iterations in 3 chains, with a burn-in of 20,000 iterations and a thinning rate of 4 for our posterior estimates. We visually assessed the mixing and convergence of each of the chains, and then used Gelman-Rubin statistics (Gelman and Rubin 1992) on the annual abundance estimates to ensure each of the models converged.

**Table 1.** *Annual harvest quotas and observed harvests in each bear management zone (A, B, C & D) in Wisconsin from 2011-2016.*

|      | Quota | | | | Harvest | | | |
| --- | --- | --- | --- | --- | --- | --- | --- | --- |
| Year | A | B | C | D | A | B | C | D |
| 2011 | 3465 | 1510 | 2550 | 1480 | 1592 | 969 | 715 | 975 |
| 2012 | 3425 | 1335 | 2970 | 1285 | 1907 | 841 | 810 | 1082 |
| 2013 | 2130 | 690 | 4110 | 1630 | 1249 | 490 | 1029 | 1179 |
| 2014 | 2070 | 990 | 5050 | 2230 | 1315 | 738 | 1024 | 1444 |
| 2015 | 1875 | 1090 | 5490 | 2235 | 1119 | 764 | 972 | 1341 |
| 2016 | 1655 | 1195 | 6190 | 2480 | 1141 | 850 | 1067 | 1624 |

**Results**

*Summary Statistics*

The mean harvest quota across the state was 9,855 (± 476 SE) bears and ranged from 8,560 to 11,520 harvest permits. The harvest quota for zone A decreased over time, with a mean of 2,437 (± 326 SE,

range = 1,655-3,465), while the quota for zone B initially decreased and then increased with a mean of 1,135 (± 116 SE, range = 690-1,510). The harvest quotas in zones C and D increased over time, with a mean of 4,393 (± 588 SE, range = 2,550-6,190) in Zone C and a mean of 1,890 (± 199 SE, range = 1,285-2,480) in Zone D.

The observed mean harvest across the state was 4,377 (± 118 SE) bears and ranged from 3,952 to 4,682 bears. The observed harvest in Zone A decreased over time with a mean harvest of 1,387 (± 125 SE, range = 1,119-1,907). The observed harvest in Zone B decreased initially with decreases in the quota and then increased as quotas increased with a mean harvest of 775 (± 66 SE, range = 490-969). The observed harvest in Zone C increased over the first two years and then became stable with a mean harvest of 936 (± 58 SE, range = 715-1,067), while harvest in Zone D increased with a mean of 1,274 (± 99 SE, range = 975-1,624) (Table 1). Bears with known age and sex comprised, on average, 86.8% of the annual harvest in each zone (A = 89.0%, B = 87.6%, C = 84.9%, D = 85.9%) and was stable across zones. The age of harvested bears of both sexes exhibited slight decreases for each zone (Figure 1).

*Population Estimates*

We used state-space models using AAH data in a Bayesian framework to estimate the black bear population for each management zone in Wisconsin using harvest data from 2011–2016. The zone-specific black bear abundance estimates from 2011 to 2017 indicated a decreasing trend in zone A (mean = 7293, SE = 592, range = 5286-9674) and zone B (mean = 6667, SE = 244, range = 5600-7453). In contrast, the trend was generally stable in zone C (mean = 9004, SE = 151, range = 8430-9439) and zone D (mean = 9172, SE = 107, range = 8573-9479) (Figure 2). The 95% credible intervals were reasonable and generally stable, but variation increased slightly in the final two years of estimation (Figure 2). The Gelman-Rubin scores were ≤1.001 for each year in each zone, indicating model convergence for all models.

*Figure 1.* Trends in mean age of harvest for male and female bears in each management zone in Wisconsin.

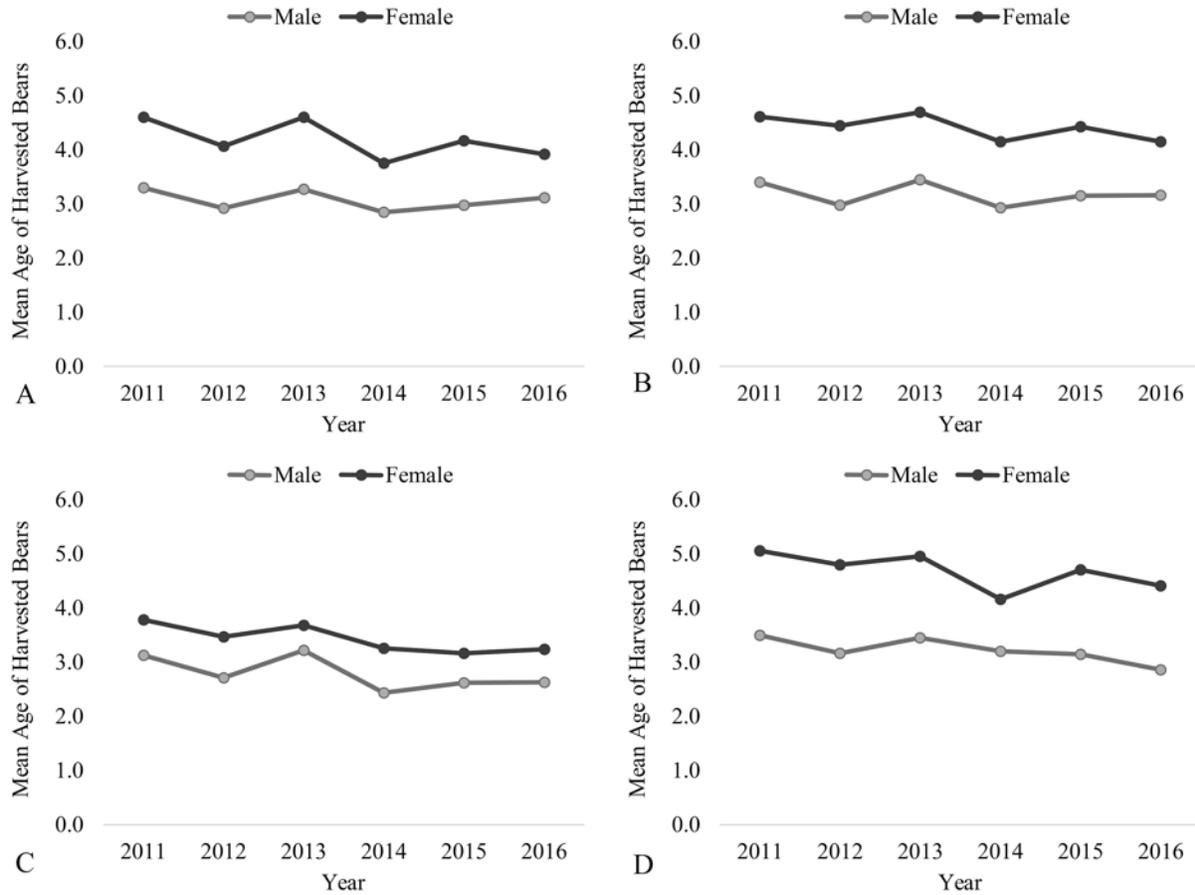

*Demographic Parameter Estimates*

    The posterior means and distributions for pregnancy rates, sex ratio of cubs or reporting rates were not updated by the model for any of the management zones. The posterior means and distributions of litter sizes were increased from the prior distributions for Zone C, but were not notably updated for the other zones. The means of the posterior distribution for the initial population size were generally similar to the means of the prior distribution for both sexes in each management zone, with the exception of 1.5 year old and 10.5+ year old age classes. The number of 1.5 year old age class males and females increased in each zone in the model, and the number of 10.5 year old age class females increased in zones B and D.

*Figure 3.* Abundance estimates of AAH state-space model population estimation for each bear management zone in Wisconsin, USA and their 95% credible intervals.

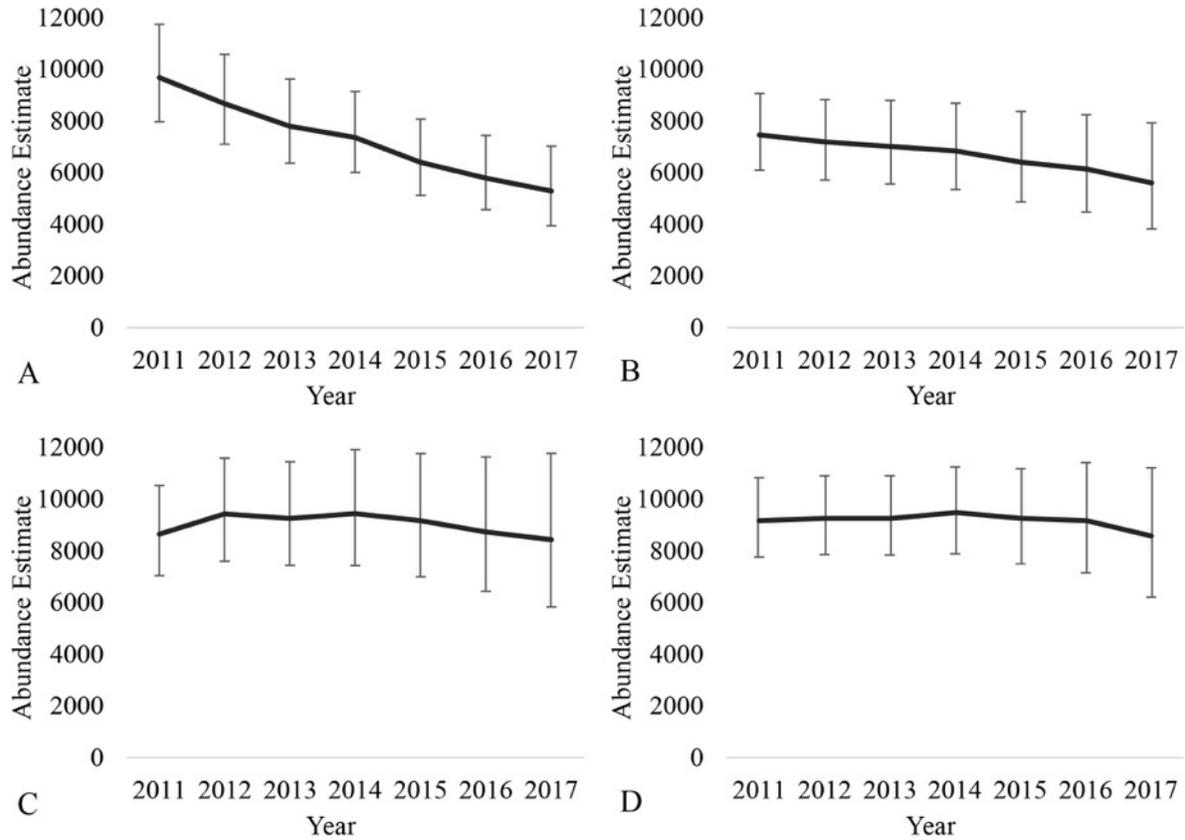

The posterior precision for harvest season survival for each sex and year in each zone were updated by the model with improved precision from the prior distributions, with the means of the posterior distributions higher or similar for zones A and D, generally higher for zones B and C. The posterior distributions for non-harvest season survival were updated by the model but had less precision and lower means and distributions for each zone than the prior distributions. The posterior distribution for cub survival had increased precision and higher means compared to the prior distribution in each zone.

**Discussion**

The posterior distribution estimates of abundance and demographic parameters from our state-space models using AAH data were reasonable for each management zone, and in many cases the estimates were updated by the model. The abundance estimates indicated a decreasing trend in zones A and B, and a generally stable trend in zones C and D. The posterior means and distributions were updated from the prior distributions for initial population size, harvest season survival, and non-harvest season survival, with a notable increase in precision for the survival values. Our posterior distributions were based on the joint modeling of the biological population process and age-at-harvest data and parameter constraints imposed via our prior specification (Norton 2015, Allen et al. 2018a). The updated posterior distributions for some of the parameters when compared to prior distributions demonstrated advantages of the state-space structure with age-at-harvest data compared to deterministic accounting-type models which do not use age-at-harvest data to inform biological parameters.

One important improvement to this model from that of Allen et al. (2018) was changing the initial population sizes of each sex and age class, which also removed the assumption of harvest being representative of availability among age classes. We adjusted the assignment of age classes to account for a bias in harvest of male bears (Malcolm and Van Deelen 2010), and also changed the initial population values from the WDNR estimates used by Allen et al. (2018) to independent population estimates from tetracycline marking for each zone (Rolley and Macfarland 2014) as our starting point. The posterior means and distributions for initial population estimates were generally similar to the prior means and distributions for all age classes, with the exception of 1.5 year old bears and 10.5+ year old females in some zones. The initial underestimate of 1.5 year old bears in our prior distributions indicates higher harvest pressure for younger bears, as is often the case for black bears throughout their range (Beston 2011). We also tried to account for the abundance of long-lived bears in the Wisconsin population by considering the 10.5+ age class to be three times the abundance of the 10.5 age class. This worked for male bears in all zones, but underestimated the females in this class in two zones, likely because older females in these zones had higher survival than in our prior distributions. More age classes could be added to the model, but would result in slower computation times. Due to these adjustments, our prior distribution more closely matched the posterior distribution from the Allen et al. (2018) model, generally allowing greater accuracy, but also likely allowing the model to converge more quickly.

After the initial round of independent population estimates from tetracycline showed that the population had been underestimated by about 2/3 (MacFarland 2009), the harvest quotas increased substantially beginning in 2009 and the number of bears harvested annually in Wisconsin are now the highest of any state in the lower 48 states. Such substantial changes in harvest quotas can have important effects on wildlife populations as well as on hunter populations and motivation (Heberlein and Kuentzel 2002, Allen et al. 2019). The black bear populations in each zone also showed slightly declining ages of harvested bears, possibly due to the increased quotas, but these do not appear to be substantially affecting the population trend to date. The varying harvest quotas over time are likely to affect annual harvest, and could be integrated into the model directly rather than inferred through the observed harvest. Hunter effort is often an integral component to population reconstruction models (Skalski et al. 2011), and could be integrated into state-space models as well.

Harvest season survival is one of the most variable demographic parameters, particularly among sex and age classes, but is also the variable most informed by the AAH data. Hunters in Wisconsin are known to have male-biased harvest selection for black bears (Malcolm and Van Deelen 2010) and other species (Allen et al. 2018b), and among bears this is likely due partly to males being larger and females with dependent cubs being illegal to harvest. Males and females of many carnivore species, including black bears, also vary in their behavior (Taylor et al. 2015, Allen et al. 2016b), which can also affect their vulnerability. Harvest season survival is one of the most easily managed parameters, because it correlates closely with annual harvest quotas, and we incorporated annual stochasticity into this parameter. The modeled stochasticity does not incorporate the reproductive state of females (with or without cubs), because we did not want to make the model too complex. Stochasticity can be incorporated into any variable if needed, however, but adding complexity slows down the time to run a given model and should be considered by managers in each specific case.

Previous AAH state-space models in a Bayesian framework were most sensitive to the prior distributions of reporting rates (Norton 2015, Allen et al. 2018a), hence rigorous estimation of reporting rates should be a focus of future research. Changes in reporting rates are likely to change in Wisconsin as the registration requirements shift from mandatory in-person registration to online registration in Wisconsin. This may result in lower reporting rates than the near universal reporting that we assumed in our analysis. Because AAH state-space models are particularly sensitive to prior distributions of reporting rate (Norton 2015, Allen et al. 2018a), conservative estimates may be best.

Fitting population models within a state-space Bayesian framework is a powerful and flexible approach that can be optimized to specific research or management objectives. The accuracy of AAH state-space models could potentially be improved by creating zone-specific estimates for prior distributions of demographic parameters based on zone-specific prior knowledge or by including auxiliary data. This could be particularly useful where it is reasonable to assume that environmental or anthropomorphic influences, such as supplemental feeding, on parameters vary between zones. Litter sizes and pregnancy rates values may also vary by zone, as they are tied to available nutrition (Rogers 1987, Noyce and Garshelis 1994). Bears in Wisconsin eat large amounts of supplemental feed (Kirby et al. 2017), but this is variable across both coarse and fine scales on the landscape, as are other food sources. Using hunter surveys to understand amounts of supplemental feed available by county or zone could inform the prior distributions set for demographic parameters.

**ACKNOWLEDGMENTS**

We thank the Wisconsin Department of Natural Resources and the Department of Forest and Wildlife Ecology at the University of Madison, and the Illinois Natural History Survey at the University of Illinois for their support. This project was funded by Federal Aid in Wildlife Restoration Grant WI W-160-R. We thank S. Hull, D. MacFarland, G. Stauffer, B. Dhuey, R. Rolley, Y. Luo, Q. Li, and J. Rees for their support.

## Literature Cited

Allen, M. L., B. Kohn, N. M. Roberts, S. M. Crimmins, and T. R. Van Deelen. 2017. Benefits and drawbacks of determining reproductive histories for black bears (*Ursus americanus*) from cementum annuli techniques. Canadian Journal of Zoology 95:991–995.

Allen, M. L., A. S. Norton, G. Stauffer, N. M. Roberts, Y. Luo, Q. Li, and T. R. Van Deelen. 2018a. A Bayesian state-space model using age- at-harvest data for estimating the population of black bears (*Ursus americanus*) in Wisconsin. Scientific Reports 8:12440.

Allen, M. L., N. M. Roberts, and T. R. Van Deelen. 2018b. Hunter selection for larger and older male bobcats affects annual harvest demography. Royal Society Open Science 5:180668.

Allen, M. L., N. M. Roberts, M. J. Farmer, and T. R. Van Deelen. 2019. Decreasing available bobcat tags appear to have increased success, interest, and participation among hunters. Human Dimensions of Wildlife 24.

Allen, M. L., C. C. Wilmers, L. M. Elbroch, J. M. Golla, and H. U. Wittmer. 2016a. The importance of motivation, weapons, and foul odors in driving encounter competition in carnivores. Ecology 97:1905–1912.

Allen, M. L., V. Yovovich, and C. C. Wilmers. 2016b. Evaluating the responses of a territorial solitary carnivore to potential mates and competitors. Scientific Reports 6:27257.

Beston, J. A. 2011. Variation in life history and demography of the American black bear. Journal of Wildlife Management 75:1588–1596.

Buckland, S. T., K. B. Newman, L. Thomas, and N. B. Koesters. 2004. State-space models for the dynamics of wild animal populations. Ecological Modelling 171:157–175.

Caswell, H. 2001. Matrix population models: construction, analysis, and interpretation. 2nd edition. Sinauer Associates.

Conn, P. B., D. R. Diefenbach, J. L. Laake, M. A. Ternent, and G. C. White. 2008. Bayesian analysis of wildlife age-at-harvest data. Biometrics 64:1170–1177.

Fieberg, J. R., K. W. Shertzer, P. B. Conn, K. V. Noyce, and D. L. Garshelis. 2010. Integrated population

modeling of black bears in minnesota: implications for monitoring and management. PLoS ONE 5.

Garshelis, D. L., and H. Hristienko. 2006. State and provincial estimates of American black bear numbers versus assessments of population trend. Ursus 17:1–7.

Gelman, A., and D. B. Rubin. 1992. Inference from iterative simulation using multiple sequences. Statistical Science 7:457–472.

Heberlein, T. A., and W. F. Kuentzel. 2002. Too many hunters or not enough deer? Human and biological determinants of hunter satisfaction and quality. Human Dimensions of Wildlife 7:229–250.

Karanth, K. U., and J. D. Nichols. 1998. Estimation of tiger densities in India using photographic captures and recaptures. Ecology 79:2852–2862.

Kirby, R., D. M. Macfarland, and J. N. Pauli. 2017. Consumption of intentional food subsidies by a hunted carnivore. Journal of Wildlife Management 81:1161–1169.

Leopold, A. 1986. Game management. University of Wisconsin Press, Madison, WI.

MacFarland, D. M. 2009. Population estimation, habitat associations and range expansion of black bears in the upper midwest. University of Wisconsin.

Malcolm, K. D., and T. R. Van Deelen. 2010. Effects of habitat and hunting framework on American black bear harvest structure in Wisconsin. Ursus 21:14–22.

Newman, K. B., C. Fernández, L. Thomas, and S. T. Buckland. 2009. Monte Carlo inference for state-space models of wild animal populations. Biometrics 65:572–583.

Norton, A. S. 2015. Integration of harvest and time-to-event data used to estimate demographic parameters for white-tailed deer. University of Wisconsin, Madison.

Noyce, K. V., and D. L. Garshelis. 1994. Body size and blood characteristics as indicators of condition and reproductive performance in black bears. Bears: Their Biology and Management 9:481–496.

Pianka, E. R. 1970. On r- and K-Selection. American Naturalist 104:592–597.

Plummer, M. 2003. JAGS: A Program for Analysis of Bayesian Graphical Models Using Gibbs Sampling. Proceedings of the 3rd International Workshop on Distributed Statistical Computing 3:1–10.

Plummer, M. 2006. rjags. R Package.

Rogers, L. L. 1987. Effects of food supply and kinship on social behavior, movements, and population growth of black bears in northeastern Minnesota. Wildlife Monographs 97:1–72.

Rolley, R. E., and D. M. Macfarland. 2014. Black Bear Population Analyses 2014.

Sadeghpour, M. H., and T. F. Ginnett. 2011. Habitat selection by female American black bears in northern Wisconsin. Ursus 22:159–166.

Skalski, J. R., J. J. Millspaugh, M. V. Clawson, J. L. Belant, D. R. Etter, B. J. Frawley, and P. D. Friedrich. 2011. Abundance trends of American martens in Michigan based on statistical population reconstruction. Journal of Wildlife Management 75:1767–1773.

Skalski, J. R., K. E. Ryding, and J. J. Millspaugh. 2005. Wildlife demography: analysis of sex, age, and count data. Elsevier Academic Press.

Taylor, A. P., M. L. Allen, and M. S. Gunther. 2015. Black bear marking behaviour at rub trees during the breeding season in northern California. Behaviour 152:1097–1111.